\documentclass[preprint,12pt,longnamesfirst]{aastex}

\begin{document}

\title{The Nature of SN\,1961V}
\shorttitle{SN\,1961V}

\author{You-Hua Chu\altaffilmark{1}, Robert A.\
  Gruendl\altaffilmark{1}, Christopher J.\ Stockdale\altaffilmark{2},
  Michael P. Rupen\altaffilmark{3}, John J.\ Cowan\altaffilmark{4}, 
  Scott W.\ Teare\altaffilmark{5}}
\altaffiltext{1}{Astronomy Department, University of Illinois, 
        1002 W. Green Street, Urbana, IL 61801;
        chu@astro.uiuc.edu, gruendl@astro.uiuc.edu}
\altaffiltext{2}{Department of Physics, Marquette University, PO Box 
        1881, Milwaukee, WI 53201-1881; christopher.stockdale@mu.edu}
\altaffiltext{3}{National Radio Astronomical Observatories, PO Box 0,
        1003 Lopezville Road, Socorro, NM 87801-0387; mrupen@nrao.edu}

\altaffiltext{4}{Department of Physics and Astronomy, University of
        Oklahoma, 440 West Brooks, Room 131, Norman, OK 73019;
        cowan@mail.nhn.ou.edu}
\altaffiltext{5}{Department of Electrical Engineering, New Mexico
        Tech, 801 Leroy Place, Socorro, NM 87801; teare@nmt.edu}


\begin{abstract}

The nature of SN\,1961V has been uncertain.  Its peculiar
optical light curve and slow expansion velocity are
similar to those of super-outbursts of luminous blue 
variables (LBVs), but its nonthermal radio spectral index 
and declining radio luminosity are consistent with 
decades-old supernovae (SNe).  
We have obtained {\it Hubble Space Telescope} STIS images and 
spectra of the stars in the vicinity of SN\,1961V, and 
find Object 7 identified
by Filippenko et al.\ to be the closest to the optical and 
radio positions of SN\,1961V.
Object 7 is the only point source detected in our STIS spectra
and only its H$\alpha$ emission is detected; it cannot be the
SN or its remnant because of the absence of forbidden lines.
While the H$\alpha$ line profile of Object 7 is remarkably
similar to that of $\eta$ Car, the blue color (similar to
an A2\,Ib supergiant) and lack of appreciable variability are 
unlike known post-outburst LBVs.
We have also obtained Very Long Baseline Array (VLBA) observations
of SN\,1961V at 18~cm.
The non-detection of SN\,1961V places a lower limit on the size
of the radio-emitting region, 7.6 mas or 0.34 pc, which implies 
an average expansion velocity in excess of 4,400 km~s$^{-1}$, 
much higher than the optical expansion velocity measured in 1961.
We conclude the following: (1) A SN occurred in the vicinity of 
SN\,1961V a few decades ago.
(2) If the SN\,1961V light maximum originates from a giant eruption 
of a massive star, Object 7 is the most probable candidate for the
survivor, but its blue color and lack of significant variability
are different from a post-outburst $\eta$ Car.
(3) The radio SN and Object 7 could be physically associated with
each other through a binary system.  
(4) Object 7 needs to be monitored to determine 
its nature and relationship to SN\,1961V.

\end{abstract}  

\keywords{supernovae: general -- supernovae: individual 
 (SN\,1961V) -- supergiants -- galaxies: individual (NGC\,1058)
 -- radio continuum}

\section{Introduction}

SN\,1961V in the outskirts of the Sc galaxy NGC\,1058 
\citep[distance = 9.3 Mpc;][]{T80,Setal96} is the prototype of 
Zwicky's Type~V supernovae (SNe), which are now classified as 
Type II Peculiar SNe.  
SN\,1961V was unusual in many respects \citep{BG71}.
First, SN\,1961V is one of the very few SNe whose progenitors
are known.
The progenitor of SN\,1961V was visible as an 18 mag star from 
1937 to 1960 \citep{Z64}.
Second, the optical light curve of SN\,1961V was more complex 
and much broader than any other SN ever observed.  
It peaked at $\sim$12 mag in late 1961, remained visible for
several years, and faded to 21.7 mag in 1970 \citep{BG71,BA70}.
Third, the initial expansion velocity of SN\,1961V, 2,000--3,000 
km s$^{-1}$, was much lower than the typical expansion velocity 
of 15,000--20,000 km s$^{-1}$ for most SNe \citep{Z64,BG71}.

These peculiar properties of SN\,1961V have raised skepticism 
concerning its nature as a SN.  
Based on the extended optical light curve and the anomalously low 
expansion velocity, \citet{Getal89} suggested that SN\,1961V was 
a luminous blue variable (LBV) similar to $\eta$~Car, and its 
``SN explosion" was actually a super-outburst. 
\citet{Fetal95} obtained {\it Hubble Space Telescope (HST)} WFC1 
images of SN\,1961V and identified a red star with $R$ = 24.55, 
their Object 6, as a candidate for the LBV survivor.
Using archival {\it HST} WFPC2 images, \citet{VDFL02} suggested
a fainter and redder star as an alternative candidate for the 
LBV survivor and called it Object 11 as an extension of the
\citet{Fetal95} object list.
The LBV hypothesis cannot explain, however, the nonthermal radio 
spectral index of SN\,1961V, $-0.4\pm0.3$ \citep{CHB88} or 
$-0.79\pm0.23$ \citep{Setal01b}.
The radio emission from $\eta$ Car and its ejecta is thermal and
optically thick, as it shows a positive spectral index and a
complete lack of polarization \citep{Detal95,DWL97}.
Furthermore, SN~1961V is more than 1,000 times more luminous than
$\eta$ Car at 20-cm wavelength \citep{R83,Setal01b} and its radio light 
curve is similar to those of radio SNe \citep{Setal01a}.

Existent images of SN\,1961V show a complex environment.
Narrow-band, emission-line images reveal two \ion{H}{2} regions
within 3$''$ to the north of SN\,1961V \citep{F85,CHB88}, while 
broad-band continuum {\it HST} WFC1 images show massive stars in 
and around the \ion{H}{2} regions and in the vicinity of 
SN\,1961V \citep{Fetal95}.
SN\,1961V is clearly associated with a star-forming environment,
where SNe are expected to occur.
Indeed, radio observations \citep{CHB88,Setal01b} show a 
nonthermal source at the south edge of the eastern \ion{H}{2} 
region, coincident to within 1$''$ of the optical position of 
SN\,1961V, and a fainter nonthermal radio source in the western
\ion{H}{2} region, corresponding to an unrelated supernova 
remnant (SNR).

To determine the nature of SN\,1961V, it is crucial to recover 
the optical counterpart of SN\,1961V with a high degree of
certainty, and to determine spectroscopically whether it is a 
surviving LBV or a SN turning into a SNR.
Thus, we have obtained {\it HST} imaging and spectroscopic
observations of SN\,1961V.
We have also obtained high-resolution radio observations to
determine the size of the radio source at SN\,1961V.
This paper reports these observations of SN\,1961V and our 
analysis of its nature.

\section{Observations and Reduction}

\subsection{HST STIS Observations}

SN\,1961V was observed with the {\it Hubble Space Telescope} 
Imaging Spectrograph (STIS) on 2002 October 23.  
Medium-resolution optical spectra were obtained with the G430M and 
G750M gratings tilted to observe the spectral ranges centered at 
4961~\AA\ and 6581~\AA, respectively.
The wavelength coverage and spectral resolution of the resulting 
spectrograms are 4800-5125~\AA\ and $\sim$0.8 \AA\ for the G430M 
granting, and 6250--6900~\AA\ and $\sim$1.3 \AA\ for the G750M 
grating.
The 52\arcsec$\times$2\arcsec\ slit was used in order to obtain
``slitless'' spectra of multiple sources simultaneously.
The long dimension of the slit was oriented along a position angle 
of 139\fdg6 to optimize the projected separations among the 
candidates of SN\,1961V's optical counterpart, minimizing the 
confusion from overlapping spectra.

The target position was acquired by peaking up on a nearby star 
and then offsetting to the location of SN\,1961V.
Spectrograms were then obtained using CR-SPLIT$=$4 for a total 
exposure time of 4,377 and 3,902 s for the G750M and G430M 
gratings, respectively.
In between the spectroscopic observations, a deep image of the 
field with the 50CCD aperture and no filter was obtained for 
comparison with the spectroscopic observations.  This imaging 
observation also used a CR-SPLIT$=$4 and had a total integration 
time of 1,800 s.  

The raw observations were reduced using the STSDAS v3.0 and 
CALSTIS v2.3 utilities within IRAF.  For this reduction the 
best reference files were obtained from the Multimission 
Archive at Space Telescope.  
The sources in the field are too faint for the observations 
to be dithered to aid in the rejection of hot/warm pixels in 
the STIS/CCD images/spectra; therefore, the standard reduction 
pipeline was used with some attempt made to construct a better 
dark frame than was available.  
The raw dark frames obtained as part of the STIS calibration 
programs 9605 and 9612, which occurred closest in time to our 
STIS observations, were uploaded and combined.  We found that 
even restricting the darks used to those from the five days 
bracketing our observations did not improve our image quality.
We also investigated making changes to the standard pipeline 
processing where the dark frame was scaled to account for the 
difference in exposure time and CCD housing temperature.
We found that a slight improvement to the image quality 
could be made by reducing the pipeline's default scaling 
of the darks by 10\% for our imaging data and 2\% for our 
spectroscopic data.
The effect of these changes is to minimize the excursions of 
improperly subtracted hot pixels from the median value in the 
frame.

An astrometric solution in the ICRS 2000.0 frame for the 
STIS/CCD image of the region around SN\,1961V was obtained by 
first finding an astrometric solution for a WFPC2 image taken 
with the F606W filter as part of program 5446.
This astrometric solution for the WFPC2 image is based on the 
positions of 7 stars with low proper motions from the USNO B1.0 
catalog and has an accuracy of better than 0\farcs2 based on 
the residuals of the least squares fit to the stellar positions.
We then used 19 point sources that appear in both the WFPC2 and 
STIS images to obtain a bootstrapped astrometric solution for 
the STIS image.  
The accuracy of the astrometry for the STIS image is estimated 
to be $\sim$0\farcs25 based on the rms of the residuals from the 
least squares fit bootstrap solution added in quadrature to the
uncertainty for the astrometric solution for the WFPC2 image.

\subsection{VLBA Observations}

We have made high-resolution radio observations of SN~1961V 
using the Very Long Baseline Array (VLBA)\footnote{The 
National Radio Astronomy Observatory is a facility of the 
National Science Foundation operated under a coorperative 
agreement by Associated Universities, Inc.}.  
The observations were made at 18~cm on 1999 September 14.
3C286 and 3C48 were used for flux calibration and J0253+3835 
was used for phase calibration \citep{PBZ03}.
SN~1961V is a weak source for the VLBA \citep{CHB88}, so the 
phased Very Large Array (VLA) in its A-configuration, the longest
baseline mode, was included to improve the sensitivity.
The 27 VLA antennae were phased and sampled every 10 seconds, 
and this sampling rate was used for the Very Long Baseline 
Interferometry (VLBI) data set.
The data set was averaged in frequency to a total of 8 
intermediate frequencies (IF's).
We iteratively self-calibrated the VLBA observations of 
J0253+3835 for a single IF to produce a common amplitude 
scale, and this model was applied to the other IF's to produce 
a VLBA phase solution.
The VLBA and VLA observations of J0253+3835 were then combined
and self-calibrated further to achieve the best possible VLBI 
phase solutions. 
For a consistency check of our phase solution, embedded scans 
of a secondary calibrator, J0230+4032, were obtained throughout 
the course of the observations of SN\,1961V.

The VLBI phase solution was applied to the observations and the 
resulting visibilities were Fourier transformed to obtain an 
image of SN\,1961V.
The synthesized beam size for this VLBI image was 11.5 mas 
$\times$ 7.56 mas, and the rms noise in the map was 0.030 mJy 
beam$^{-1}$.
No compact or diffuse sources were detected with $\ge 3 \sigma$ 
confidence within 100 mas of the radio position of SN\,1961V, 
at which the simultaneous VLA observations detected 
$0.147\pm 0.026$ mJy beam$^{-1}$ for a synthesized beam of 
1\farcs20 $\times$ 1\farcs01 \citep{Setal01b}.
To reconcile the VLA detection and VLBI non-detection, the radio 
emission must be distributed over an area greater than the VLBI 
synthesized beam, i.e., $>$ 7.56 mas.

\section{Discussion}

\subsection{Stellar and Interstellar Environment of SN\,1961V}

The broad-band STIS image in Figure 1 represents the deepest
high-resolution image of SN\,1961V and its surroundings ever
obtained.
In addition to the eleven bright stellar objects identified by 
\citet{Fetal95} and \citet{VDFL02}, many fainter stars are also 
detected.
These eleven bright objects, marked in Figure 1c, are mostly 
supergiants with $V = $ 24 to 25.5.
To illustrate the position of SN\,1961V, radio contours extracted
from the \citet{Setal01b} 18 cm  VLA map are plotted over the 
STIS image in Figure 1d.
The eastern radio source corresponds to SN\,1961V, and our new 
astrometric calibration (\S 2.1) now places Object 7 the 
closest to SN\,1961V.
Three nebular arcs are visible and form an apparent supershell 
structure about $5\farcs3 \times 3\farcs0$ (or 240 pc $\times$ 
135 pc) in size.
Encompassed within the supershell are SN\,1961V and Objects 6, 
7, 9, and 11, while projected along the supershell rim are 
Objects 5, 8, and 10.

To date, the best narrow-band images of SN\,1961V are still those
obtained by \citet{F85} using the Kitt Peak 2.1 m telescope.
His red continuum image had a limiting magnitude of $\sim$
21 mag, so it did not detect any of the stars in the vicinity of
SN\,1961V shown in our STIS image.
Two \ion{H}{2} regions are detected in the [\ion{O}{3}] image,
but they appear better resolved in the H$\alpha$+[\ion{N}{2}] 
image: the eastern \ion{H}{2} region consists of a bright 
diffuse emission region and two faint emission knots to the 
south, while the western \ion{H}{2} region is resolved into 
two emission patches with similar brightnesses.

Comparisons between our STIS image and Fesen's narrow-band
images show the following correspondences.
The bright diffuse emission from the eastern \ion{H}{2} 
region is centered on the bright Object 8 projected against 
the northeast rim of the supershell, and the two 
H$\alpha$-emission knots in the southern extension are 
coincident with Object 7 inside the supershell and  
Object 5 along the south rim of the supershell, respectively.
In the western \ion{H}{2} region, the eastern emission patch 
is coincident with the bright Object 10, but the western 
emission patch has no detectable stellar counterparts in our 
STIS image.
The image of Object 10 is elongated and somewhat resolved, 
indicating that it consists of multiple stars; its location
inside an \ion{H}{2} region further indicates that it contains
a group of massive stars.
The stars in Object 10 are probably at least a few Myr old,
as they are coincident with a radio SNR identified by 
\citet{Setal01b}, the western radio source in Figure~1d.

In summary, the STIS image illustrates that SN\,1961V is 
associated with a complex, extended star forming region.  
It is projected within the central cavity of a supershell, 
with bright \ion{H}{2} regions distributed along the 
shell rim.
The presence of both \ion{H}{2} regions and SNRs indicates
that a mixture of stellar populations at different ages
exist within $\sim$100 pc of SN\,1961V.

\subsection{The Optical Counterpart of SN 1961V}

The optical position of SN\,1961V was determined by 
\citet{K86} to an accuracy better than 0\farcs1, and
this position is coincident with a fading nonthermal 
radio source \citep{BC85,CHB88,Setal01b}.
As shown in Figure~1d, Object 7 is the closest to 
SN\,1961V, the eastern radio source.

Our long-slit STIS spectroscopic observations have been
designed to detect line emission, as the stellar sources 
are faint and the detection of their continuum requires 
unrealistically long exposures.
Of all stellar sources in the slit, only one is detected
in H$\alpha$ emission (see Fig.~2).
Aligning the STIS direct image with the H$\alpha$ 
spectrogram, we find that the point source corresponds 
to Object 7.
The wavelength scale in Figure~2 is calibrated for the 
slit center; as Object 7 is offset by 0\farcs46 from the
slit center, a wavelength correction of $-$5.1 \AA\ should
be applied.\footnote{The STIS spectrogram has a spectral scale
of 0.554 \AA~pixel$^{-1}$ and a spatial scale of 
0\farcs05 pixel$^{-1}$; therefore, a 0\farcs46 offset 
from the slit center corresponds to a wavelength offset 
of $0.554 \times 0.46 / 0.05$ = 5.1 \AA.}
The corrected wavelength of the H$\alpha$ emission from 
Object 7 corresponds to a heliocentric velocity of 
$V_{\rm hel} = 520 \pm 10$ km~s$^{-1}$, which is consistent 
with the systemic velocity of the \ion{H}{2} regions near
SN\,1961V measured by \citet{Getal02}.
The H$\alpha$ line profile of Object 7, shown in Figure 3,
appears to have a narrow core and broad wings, which can be 
measured up to $\pm$550 km~s$^{-1}$ (limited by the noise 
level of the spectrum).
The absence of forbidden lines such as [\ion{O}{1}]
$\lambda$6300 and [\ion{O}{3}] $\lambda$5007 indicates
that the H$\alpha$ emission is stellar as opposed to 
originating from a SN/SNR transition object like SN\,1979C 
or SN\,1980K \citep{Fetal99}.

Our STIS spectrograms also detected a patch of diffuse
emission in the H$\alpha$ and [\ion{O}{3}] 
$\lambda\lambda$4959, 5007 lines.
This emission originates from the northwest rim of the 
supershell and corresponds to the northern part of the 
western \ion{H}{2} region identified by \citet{F85}.
The velocity of this diffuse emission cannot be accurately
determined, but is consistent with $V_{\rm hel} \sim
520$ km~s$^{-1}$.

\subsection{Radio Properties of SN\,1961V}

Recent VLA observations of SN\,1961V \citep{Setal01b} show
that it is a nonthermal radio source with a spectral index 
of $\alpha =-0.79\pm0.23$ ($S_{\nu}\propto \nu^{\alpha}$)
and a time decay index of $\beta =-0.69\pm0.23$ at 
$\sim$20~cm ($S_{\nu}\propto t^{\beta}$).
The spectral and temporal properties of SN\,1961V are
consistent with other decades-old radio SNe, for example,
SNe 1970G, 1957D, and 1950B \citep{Setal01a,S01}.
These properties are not consistent with radio observations 
of LBVs, which have an overall flat radio spectrum when 
convolved to the VLA resolution limit of SN~1961V, for
example, AG~Car, He~3-519, HR~Car, WRA~751, P~Cyg, and 
$\eta$~Car \citep{DW02,DW03,R83,DWL97}.

A more apt radio comparison may be made with SN\,1954J in 
NGC\,2403 \citep[3.2 Mpc;][]{Fetal01}, which has been 
established to be an outburst of the LBV ``Variable 12'', 
misidentified as a SN \citep{HD94,SHG01}.
The region near SN\,1954J was observed with the VLA at 20~cm 
about 31 yr following its outburst, but no source was 
detected with a $3\sigma$ limit of 0.35 mJy beam$^{-1}$
\citep{ECB02}.
On the other hand, the VLA detection of SN\,1961V at 18~cm
about 38 yr after the explosion indicates that it would 
been detected with a flux density of 0.98 mJy beam$^{-1}$ 
were it in NGC\,2403.
Clearly, SN\,1961V is more than three times more luminous than
SN\,1954J at 18-20~cm.

Our recent VLBI experiment (\S2.2) allows us to make a more 
definitive radio analysis than previously possible with 
resolution-limited VLA studies.
The VLBI non-detection of SN\,1961V indicates that its radio
emitting region is larger than the VLBI resolution element;
thus the minimum diameter of SN\,1961V is 7.56 mas, or 0.34 pc 
for a distance of 9.3 Mpc.
Assuming that the nonthermal radio emission detected by the VLA
observations originates from synchrotron radiation associated 
with SN shocks \citep{C82,CF94}, this minimum size implies an 
average expansion velocity in excess of 4,400 km s$^{-1}$, 
between the time of the VLBI experiment and the reported SN 
explosion, i.e., $\sim$38 yr.
This expansion velocity is much higher than the optical expansion 
velocity of 2,000 km s$^{-1}$ measured in 1961 November, toward
the end of the broad maximum in the light curve.

Such a large discrepancy in expansion velocities determined 
at optical and radio wavelengths has been seen previously in 
SN\,1986J, another Type II Peculiar SN.
The optical expansion velocity of SN\,1986J measured in 1986, 
several years after its possible explosion in 1982-1983, was
less than 1,000 km s$^{-1}$ \citep{Retal87}.
However, VLBI observations of SN\,1986J in 1990 and 1999 suggest
that its expansion velocity was 20,000 km~s$^{-1}$ at $t = $0.25
yr and 6,000 km~s$^{-1}$ at $t = $15.9 yr \citep{BBR02}, well
in excess of the optical expansion velocity.
In the case of SN\,1986J, its identification as a SN has been
commonly accepted.

\subsection{The Nature of SN\,1961V}

Was the SN\,1961V event a SN explosion or a super-outburst of 
an LBV?
The strongest support for the SN hypothesis has been provided 
by radio observations.
The nonthermal radio spectral index, high radio luminosity, and
the temporal decline of radio luminosity are all consistent with
the existence of a decades-old SN.
Furthermore, these radio properties are not like those of any 
known LBV: radio emission from LBVs are predominantly thermal in
origin, and LBVs (including SN\,1954J) are much fainter than 
SN\,1961V.

The size of SN\,1961V's emitting region may further constrain 
its nature.
Object 7 is unresolved in the {\it HST} STIS images, placing 
an upper limit of 0\farcs05 = 2.25 pc on its size.
The VLBI non-detection places a lower limit of 7.6 mas = 0.34 pc
on the diameter of the radio-emitting region.
These sizes are significantly larger than the size of
$\eta$ Car's nebula in 1995, 0.09 pc $\times$ 0.17 pc,
$\sim$150 yr after it was ejected during ``The Great Eruption'' 
in 1837-1860 \citep{HD94,SG98}.
Whether SN\,1961V was a SN explosion or an LBV outburst, the
ejected material expands and the expansion velocity can be 
estimated from the size and time lapse.
The above lower and upper limits on size require that the
expansion velocity is at least 4,000 km~s$^{-1}$ but at
most 27,000 km~s$^{-1}$.
Such high expansion velocity is consistent with a SN explosion,
but not an LBV outburst.

The hypothesis that SN\,1961V was an LBV outburst can be
verified only if the LBV survivor can be convincingly 
identified.
Our STIS observations suggest that Object 7 is the most
likely candidate for the LBV survivor because it is the 
closest to the optical and radio position of SN\,1961V and
is the only stellar object with the H$\alpha$ emission line 
detected.
However, Object 7 is not as red as $\eta$ Car after a
super-outburst.
Using {\it HST} WFC1 images taken in 1991 December, 
\citet{Fetal95} reported $V = 24.22$, $R = 23.37$, and 
$I = 23.79$ (errors $<$ 0.2 mag) for Object 7.
Using {\it HST} WFPC2 $F606W$ images taken in 1994 
September and $F450W$ and $F814W$ images taken in 2002 July, 
\citet{VDFL02} reported $B = 24.04$, $V = 23.85$, and 
$I = 23.83$ for Object 7 with errors $\sim$ 0.14 mag.
These can be compared to the LBV Variable 12 that was 
responsible for SN\,1954J:
$B = 22.7$, $V = 21.9$, $R = 21.1$, and $I = 20.9$ (errors
$\sim$ 0.2 mag) in 1999 February \citep{SHG01}.
It is clear that the color of Object 7, $B-I = 0.21$ at 
about 41 yr after the light maximum, is not as red as
the color of Variable 12, $B-I = 1.8$ at about 45 yr after
the outburst.
Because of the color difference, Object 7 at 41 yr after 
is about 1 mag brighter in $B$ but 0.6 mag fainter in $I$
compared to Variable 12 at 45 yr after the outburst.

We can also compare the H$\alpha$ line profile of Object 
7 to that of $\eta$ Car.
At the distance of SN\,1961V, 9.3 Mpc, $\eta$ Car would
not be resolved from its ejecta; therefore, we have extracted
an integrated spectrum of $\eta$ Car and its surrounding
ejecta nebula.\footnote{The echelle spectrograph on the 4m 
telescope at the Cerro Tololo Inter-American Observatory 
was used to map the kinematics of the ejecta nebula of 
$\eta$ Car in the H$\alpha$ and [\ion{N}{2}] lines in 
1996 January by Chu et al.  The spectral resolution, 
determined from the FWHM of the sky lines was $\sim$ 
12 km~s$^{-1}$. Parts of the data were published
by \citet{WDC99}.}
The ejecta nebula of $\eta$ Car shows pronounced [\ion{N}{2}]
lines relative to the H$\alpha$ line, but the integrated
spectrum of the star and the ejecta nebula is dominated by 
the stellar H$\alpha$ emission and the nebular [\ion{N}{2}] 
lines become negligible.
Figure 3 shows that the H$\alpha$ line profile of Object 7
is remarkably similar to that of $\eta$ Car.
The lack of a [\ion{N}{2}] counterpart to the H$\alpha$ emission
from Object 7 indicates that the emitting material is dense
and must be stellar, as is the case for LBVs or mass-losing 
stars in general.
Unfortunately, the previous observations of the H$\alpha$ line 
of SN\,1961V  made by \citet{Getal89} in 1986 were of much lower 
spectral and spatial resolution, so that the H$\alpha$ and 
[\ion{N}{2}] lines were blended and heavily contaminated by
bright \ion{H}{2} region emission.
It is impossible to determine whether the H$\alpha$ line width 
evolved from 1986 to present.

Is Object 7 an LBV?  
While its H$\alpha$ line profile resembles that of $\eta$ Car,
its color is not red. 
In fact, based on its colors and magnitudes \citet{Fetal95} 
assigned a spectral type of A2\,Ib to Object 7.
It is interesting to note that there exist A-type supergiants
with similar H$\alpha$ emission line profiles.
For example, the A5\,Ia-O star B324 in M33 shows an H$\alpha$
emission line with a narrow core and broad wings, but M33 B324
is not known to exhibit variability and cannot be classified
as an LBV \citep{HMF90,Hetal94}.
Similarly, the {\it HST} photometric measurements of Object 7
by \citet{Fetal95} and by \citet{VDFL02} do not show significant
variations from 1991 to 2002.
The lack of large-amplitude variability from 30 yr to 40 yr after
the light maximum is in sharp contrast with the variability of 
$\eta$ Car after its Great Eruption.
Therefore, even if Object 7 is an LBV in a dormant state,
its blue color and lack of variability are unlike known LBVs 
that have gone through super-outbursts, such as $\eta$ Car.

What is the nature of the SN\,1961V event?  
The radio observations indicate the existence of a decades-old SN.
It is possible that the SN exploded in 1961 and was responsible \
for the light maximum of SN\,1961V.
It is also possible that the SN exploded prior to 1961 and 
did not contribute directly to the light curve of SN\,1961V.
In the latter case, the light maximum of SN\,1961V might be 
caused by a catastrophic but not fatally-explosive event of 
a massive star, then the spatial coincidence would make 
Object 7 the most probable candidate for the survivor, but 
the event cannot have been a super-outburst from an $\eta$ 
Car-like LBV.
As the radio SN and Object 7 are both coincident with SN\,1961V
(well within 20 pc), there is a non-negligible possibility that 
all three objects are related to one another through a binary
system similar to the situation in SN\,1993J, where a massive
binary companion of the SN progenitor is recently identified
\citep{Metal04}.
We speculate that Object 7 could be a massive binary companion
of the radio SN's progenitor.
Either the SN itself or the SN ejecta impact on Object 7 may
be responsible for the light curve of SN\,1961V.
The impact of SN ejecta on Object 7 injects energy into its 
atmosphere and causes it to expand and form the broad H$\alpha$ 
emission line.
Similar interactions have been suggested for the binary companions
of Type Ia SNe \citep[e.g.,][]{MBF00}.

We have learned from SN\,1961V that the late evolution of massive
stars is complex and confusing.
Numerous parameters can affect the appearance of a SN or an
outburst.  
Systematic studies of a large number of peculiar SNe are needed.
In the meantime, it is necessary to monitor Object 7 photometrically
and spectroscopically in the future for variability in order to
achieve a better understanding of its nature and relationship with
SN\,1961V.

\section{Summary}

We have obtained {\it Hubble Space Telescope} STIS images and 
spectra of the stars in the vicinity of SN\,1961V.
The STIS image shows that SN\,1961V and a number of massive
stars are inside a supershell with two \ion{H}{2} regions
located on the shell rim.
Using a new astrometric solution, we find Object 7 of
\citet{Fetal95} to be the closest to SN\,1961V.
Furthermore, Object 7 is the only point source detected in 
our STIS spectra, and only its H$\alpha$ emission is detected.
The H$\alpha$ line profile of Object 7 is remarkably similar 
to that of $\eta$ Car, but its blue color (similar to
an A2\,Ib supergiant) and lack of significant variability are 
unlike any known LBVs at 30-40 yr after a super-outburst.

The nonthermal spectral index and declining radio luminosity of 
SN 1961V are consistent with those of decades-old SNe.
We have obtained VLBA observations of SN\,1961V at 18~cm;
the non-detection of SN\,1961V places a lower limit on the size
of the radio-emitting region, 7.6 mas or 0.34 pc.
This implies an average expansion velocity in excess of 
4,400 km~s$^{-1}$, much higher than the optical expansion velocity 
measured in 1961.
Such discrepant expansion velocities have been observed in other
SNe, such as SN\,1986J.

We conclude that a SN has occurred in the vicinity of SN\,1961V. 
The physical nature of SN\,1961V's light curve remains uncertain.
If it was associated with a great eruption of a massive star,
Object 7 is the most probable survivor, but its color and 
lack of significant variability differ from a post-outburst 
$\eta$ Car-type LBV.
It is possible that the radio SN and Object 7 are both associated
with SN\,1961V through a binary system.
In the future, Object 7 should be monitored photometrically
and spectroscopically for variability in order to understand
its nature and relationship to SN\,1961V.

\acknowledgments 
Support for this work was provided by NASA through grant
numbers HST-GO-09371.01-A and HST-GO-09371.02-A from the 
Space Telescope Science Institute, which is operated by the 
Association of Universities for Research in Astronomy, Inc., 
under NASA contract NAS5-26555.  
Support for this work was also provided in part by NSF, at 
the University of Oklahoma (AST-0307279).

\clearpage

\begin{figure}
\figurenum{1}
\centerline{\plotone{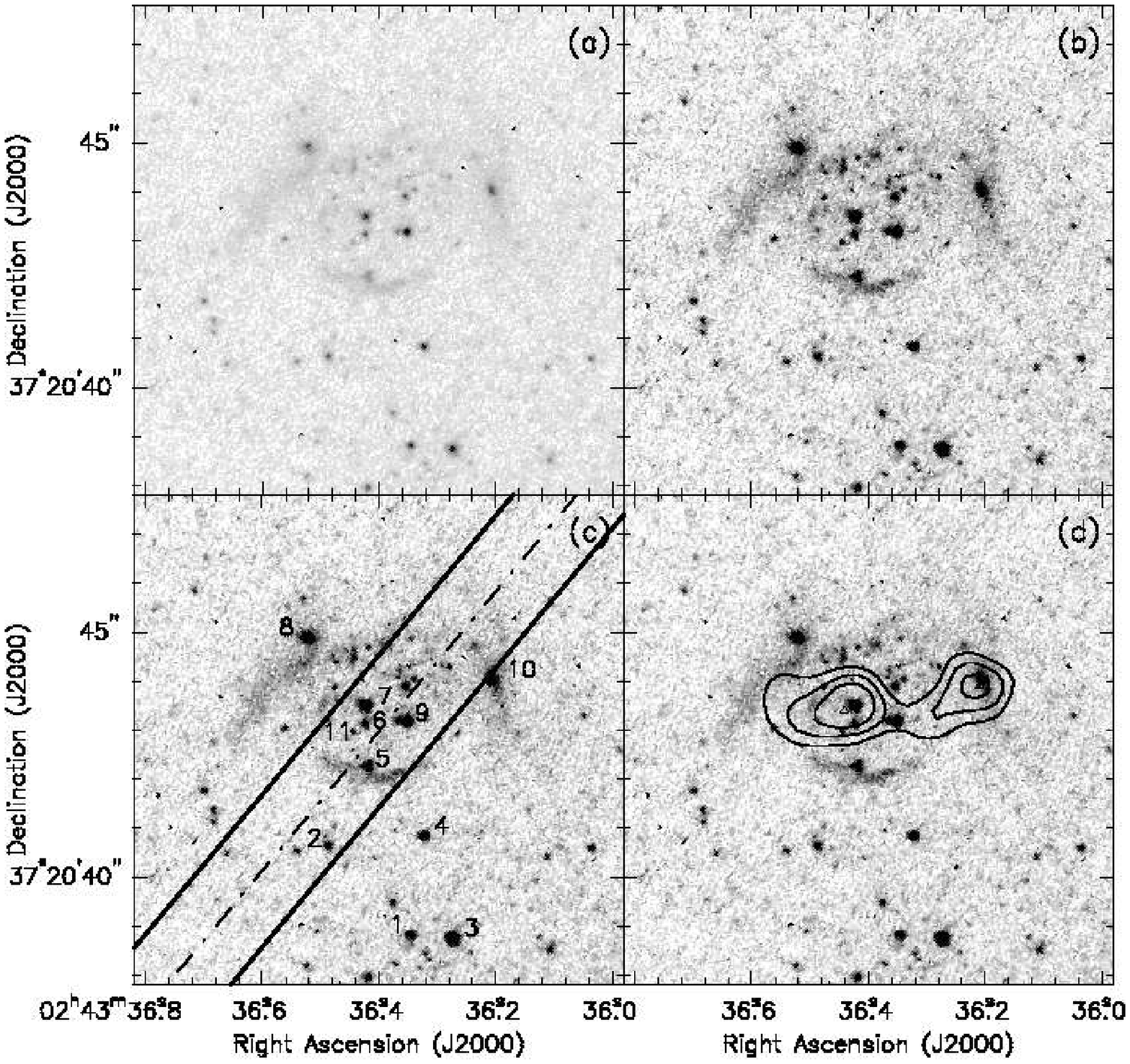}}
\caption{{\it HST} STIS image of the region around SN\,1961V.
The images are presented in two different greyscales in (a) and
(b).  The objects identified by \citet{Fetal95} and
\citet{VDFL02} are marked in (c).  The slit position for the
STIS spectroscopic observations are also marked in (c).
The contours of the VLA 18-cm map observed on 1999 September 14
\citep[Fig.~1b of][]{Setal01b} are plotted over the STIS image 
in (d).}
\label{fig1}
\end{figure}

\begin{figure}
\figurenum{2}
\epsscale{0.6}
\centerline{\plotone{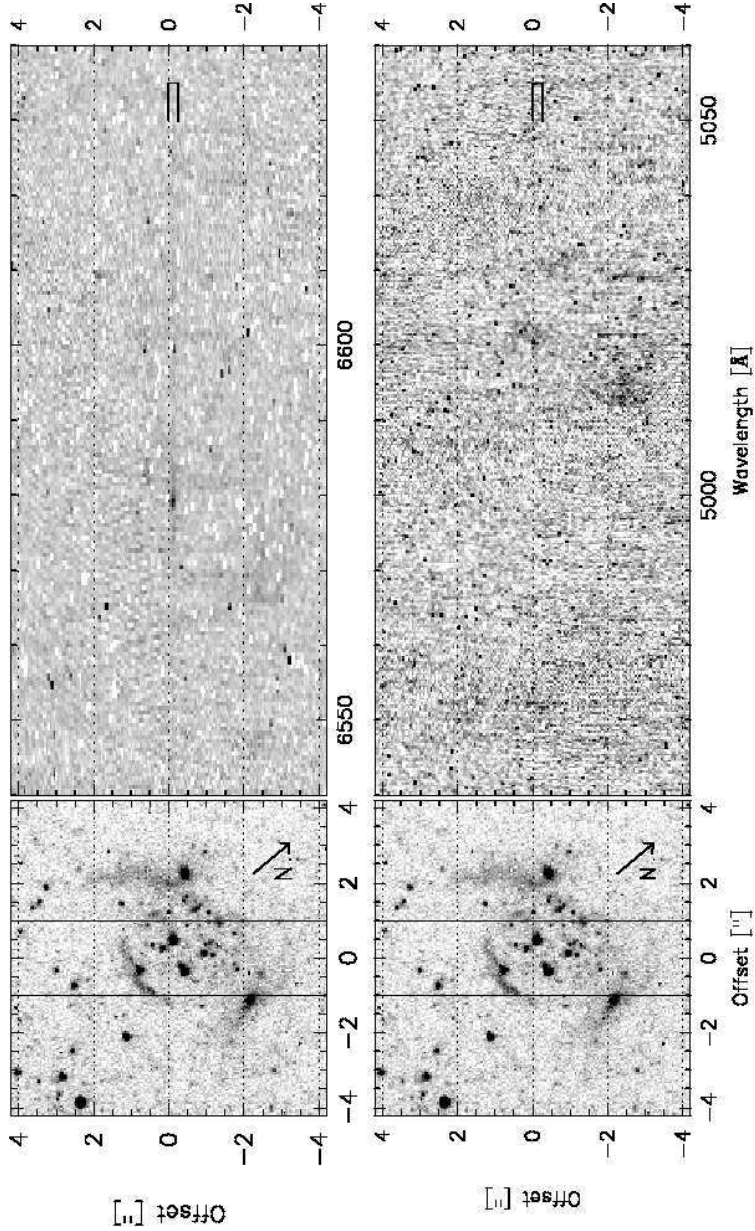}}
\caption{{\it HST} STIS image (left panels) and spectra (right
panels) of objects around SN\,1961V.  The direction of north
and the 2$''$-wide slit are marked on the STIS image.  
The upper right panel shows the spectrum around the H$\alpha$ 
line and the lower right panel the [\ion{O}{3}] lines.  
Object 7 is the only point source detected, and only
its H$\alpha$ emission line is detected.
The square bracket marks the slit extent over which the 
spectrum of Object 7 shown in Figure 3 is extracted.
A patch of diffuse emission at slit offset of $-$2$''$ to $-$3$''$ 
is detected in both the H$\alpha$ and the [\ion{O}{3}] lines.
Horizontal dashed lines are drawn to assist in the visual 
alignment of image and spectral features.
}
\label{fig2}
\end{figure}

\begin{figure}
\figurenum{3}
\epsscale{1}
\centerline{\plotone{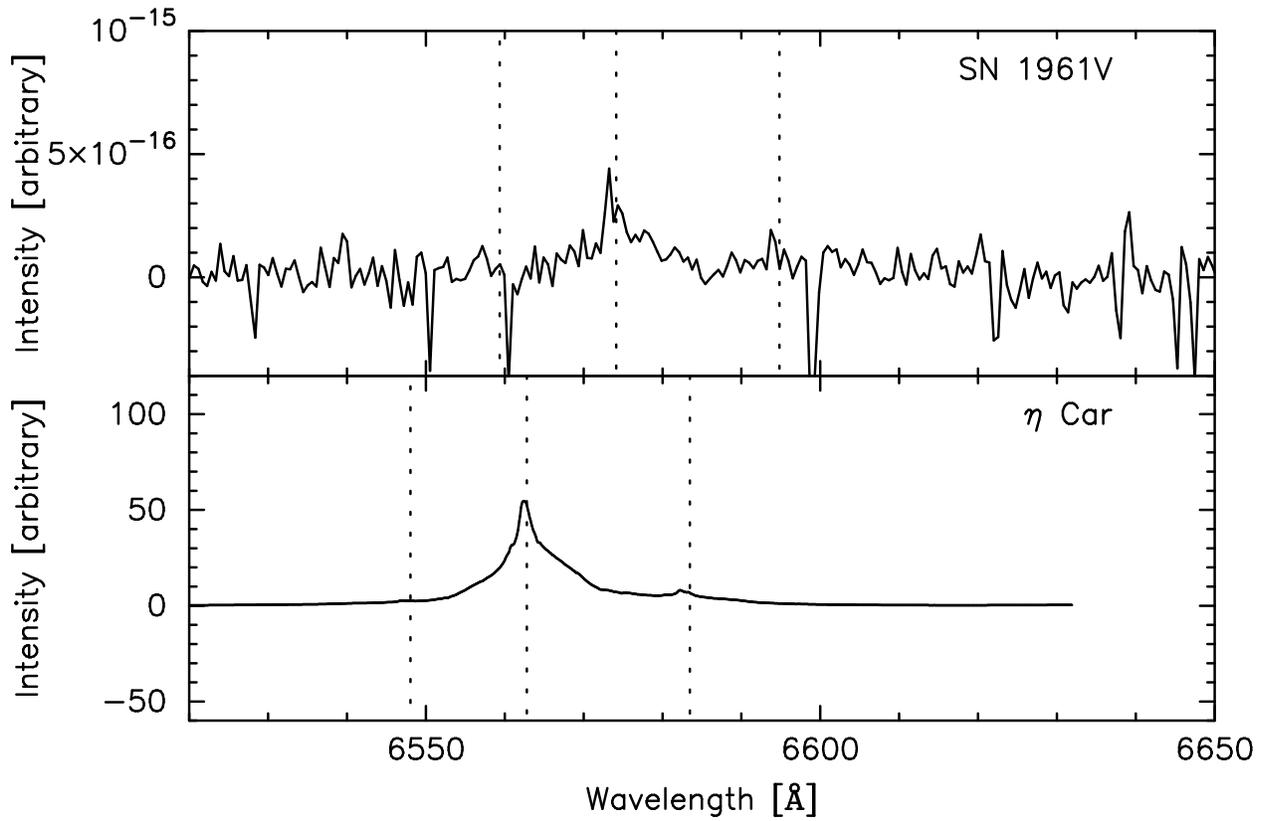}}
\caption{(a) H$\alpha$ line profile of Object 7 extracted from
our {\it HST} STIS observations.  (b) H$\alpha$ line profile
of $\eta$ Car and its surrounding ejecta nebula taken with the
echelle spectrograph on the CTIO 4m telescope.  The vertical 
dashed lines mark the locations of the H$\alpha$ and 
[\ion{N}{2}] $\lambda\lambda$6548, 6583 lines.
}
\label{fig3}
\end{figure}


\begin{thebibliography}{}

\bibitem[Bertola \& Arp(1970)]{BA70}
Bertola, F., \& Arp, H.\ 1970, \pasp, 82, 894

\bibitem[Bietenholz et al.(2002)Bietenholz, Bartel, \& Rupen]
{BBR02} Bietenholz, M.~F., Bartel, N., \& Rupen, M.~P.\ 2002, 
\apj, 581, 1132 

\bibitem[Branch \& Cowan(1985)]{BC85}
Branch, D., \& Cowan, J.~J.\ 1985, \apjl, 297, L33

\bibitem[Branch \& Greenstein(1971)]{BG71}
Branch, D., \& Greenstein, J.~L.\ 1971, \apj, 167, 89

\bibitem[Chevalier(1982)]{C82} Chevalier, R.~A.\ 1982, 
\apj, 259, 302 

\bibitem[Chevalier \& Fransson(1994)]{CF94} Chevalier, 
R.~A.~\& Fransson, C.\ 1994, \apj, 420, 268 

\bibitem[Cowan et al.(1988)Cowan, Henry, \& Branch]{CHB88}
Cowan, J.~J., Henry, R.~B.~C., \& Branch, D.\ 1988, \apj, 329, 116

\bibitem[Cowan et al.(1994)Cowan, Roberts, \& Branch]{CRB94} Cowan, 
J.~J., Roberts, D.~A., \& Branch, D.\ 1994, \apj, 434, 128 

\bibitem[Duncan \& White(2002)]{DW02} Duncan, R.~A.~\& 
White, S.~M.\ 2002, \mnras, 330, 63 

\bibitem[Duncan \& White(2003)]{DW03} Duncan, R.~A.~\& 
White, S.~M.\ 2003, \mnras, 338, 425 

\bibitem[Duncan et al.(1997)Duncan, White, \& Lim]{DWL97}
Duncan, R.~A., White, S.~M., \& Lim, J.\ 1997, \mnras, 290, 680

\bibitem[Duncan et al.(1995)]{Detal95} Duncan, R.~A., White, 
S.~M., Lim, J., Nelson, G.~J., Drake, S.~A., \& Kundu, M.~R.\ 1995, 
\apjl, 441, L73 

\bibitem[Eck et al.(2002)Eck, Cowan, \& Branch]{ECB02} Eck, C.~R., 
Cowan, J.~J., \& Branch, D.\ 2002, \apj, 573, 306 

\bibitem[Fesen(1985)]{F85} Fesen, R.~A.\ 1985, \apjl, 297, L29 

\bibitem[Fesen et al.(1999)]{Fetal99} Fesen, R.~A.~et al.\ 
  1999, \aj, 117, 725 

\bibitem[Filippenko et al.(1995)]{Fetal95} Filippenko, A.~V., 
Barth, A.~J., Bower, G.~C., Ho, L.~C., Stringfellow, G.~S., Goodrich, 
R.~W., \& Porter, A.~C.\ 1995, \aj, 110, 2261 

\bibitem[Freedman et al.(2001)]{Fetal01} Freedman, W.~L.~et al.\ 
  2001, \apj, 553, 47 

\bibitem[Goodrich et al.(1989)]{Getal89}
Goodrich, R.~W., Stringfellow, G.~S., Penrod, G.~D., \& Filippenko, 
A.~V.\ 1989, \apj, 342, 908 

\bibitem[Gruendl et al.(2002)]{Getal02} 
Gruendl, R.~A., Chu, Y.-H., Van Dyk, S.~D., \& Stockdale, 
 C.~J.\ 2002, \aj, 123, 2847 

\bibitem[Herrero et al.(1994)]{Hetal94} Herrero, A., Lennon, D.~J.,
Vilchez, J.~M., Kudritzki, R.~P., \& Humphreys, R.~H.\ 1994, \aap, 
287, 885 

\bibitem[Humphreys \& Davidson(1994)]{HD94}
Humphreys, R.~M., \& Davidson, K.\ 1994, \pasp, 106, 1025 

\bibitem[Humphreys et al.(1990)Humphreys, Massey, \& Freedman]{HMF90} 
Humphreys, R.~M., Massey, P., \& Freedman, W.~L.\ 1990, \aj, 99, 84 

\bibitem[Klemola(1986)]{K86}
Klemola, A.~R.\ 1986, \pasp, 98, 464

\bibitem[Marietta et al.(2000)Marietta, Burrows, \& Fryxell]{MBF00} 
Marietta, E., Burrows, A., \& Fryxell, B.\ 2000, \apjs, 128, 615 

\bibitem[Maund et al.(2004)]{Metal04} 
Maund, J.\ R., Smartt, S.\ J., Kudritzki, R.\ P., Podsiadlowski,
P., \& Gilmore, G.\ F.\ 2004, Nature, 427, 129

\bibitem[Perley et al.(2004)Perley, Butler, \& Zijlstra]{PBZ03}
Perley, R., Butler, B.~J. \& Zijlstra, A.\ 2004, in preparation

\bibitem[Retallack(1983)]{R83} 
Retallack, D.~S. 1983, \mnras, 204, 669

\bibitem[Rupen et al.(1987)]{Retal87} Rupen, M.~P., van Gorkom, 
J.~H., Knapp, G.~R., Gunn, J.~E., \& Schneider, D.~P.\ 1987, 
\aj, 94, 61 

\bibitem[Silbermann et al.(1996)]{Setal96}
Silbermann, N.~A., et al.\ 1996, \apj, 470, 1 

\bibitem[Smith \& Gehrz(1998)]{SG98} Smith, N.~\& Gehrz, 
 R.~D.\ 1998, \aj, 116, 823 

\bibitem[Smith et al.(2001)Smith, Humphreys, \& Gehrz]{SHG01} 
Smith, N., Humphreys, R.~M., \& Gehrz, R.~D.\ 2001, \pasp, 113, 692 

\bibitem[Stockdale(2001)]{S01}
Stockdale, C.~J.\ 2001, PhD thesis, University of Oklahoma

\bibitem[Stockdale et al.(2001a)]{Setal01a} 
Stockdale, C.~J., Goss, W.~M., Cowan, J.~J., \& Sramek, R.~A.\ 
2001a, \apjl, 559, L139 

\bibitem[Stockdale et al.(2001b)]{Setal01b} Stockdale, C.~J., 
Rupen, M.~P., Cowan, J.~J., Chu, Y.-H., \& Jones, S.~S.\ 2001b,
\aj, 122, 283 

\bibitem[Tully(1980)]{T80} Tully, R.~B.\ 1980, \apj, 237, 390

\bibitem[Van Dyk et al.(2002)Van Dyk, Filippenko, \& Li]{VDFL02} 
Van Dyk, S.~D., Filippenko, A.~V., \& Li, W.\ 2002, \pasp, 114, 700 

\bibitem[Weis et al.(1999)Weis, Duschl, \& Chu]{WDC99} Weis, K., 
 Duschl, W.~J., \& Chu, Y.-H.\ 1999, \aap, 349, 467 

\bibitem[Zwicky(1964)]{Z64} Zwicky, F.\ 1964, \apj, 139, 514


\end{thebibliography}
\end{document}